\documentclass[prb,reprint,amsmath,amssymb,showpacs,superscriptaddress,floatfix,longbibliography]{revtex4-1}
\usepackage[breaklinks=true,colorlinks,citecolor=blue,linkcolor=blue,urlcolor=blue]{hyperref}
\usepackage{floatrow}
\usepackage{bm}

\usepackage{epsfig,mathrsfs,color,latexsym,subfigure,marginnote}
\renewcommand{\BibitemShut}[1]{}

\usepackage{csquotes}

\begin{document}

\title{External Strain Induced Semi-metallic and Metallic Phase of Chlorographene}
\author{Shivam Puri}\email[]{spuri@iitk.ac.in}
\author{Somnath Bhowmick}\email[]{bsomnath@iitk.ac.in}
\affiliation{Department of Material Science and Engineering, Indian Institute of Technology, Kanpur 208016, India\\}
\date{\today}

\begin{abstract}
In order to overcome the limitations of graphene due to lack of intrinsic bandgap, it is generally functionalized with hydrogen or halogen atoms like fluorine and chlorine. Generally, such functionalization yields a moderate to high bandgap material in case of 100\% coverage, for example $\approx 1.5$ eV in graphene functionalized with chlorine atoms or chlorographene. In this paper, using \textit{ab initio} calculations, we report very interesting transformations observed in chlorographene under external strain, driving it to a state with nearly vanishing bandgap (under tensile strain) and even converting it to a metal (under compressive strain). We also show the importance of spin-orbit coupling, responsible for the few meV bandgap of chlorographene observed under high tensile strain, which would have been a gapless semi-metal otherwise.
\end{abstract}
\maketitle

The past decade has seen a tremendous growth of two dimensional (2D) materials, with numerous discoveries of atomically thin layers with fascinating properties suitable for applications in next generation electronic, optoelectronic and magnetic devices.\cite{ganesh2015} The rise of 2D materials started with the first successful isolation of a single layer of graphite, known as graphene.\cite{Novoselov666,Novoselov26072005,geim2007} Post-discovery, graphene enthralled the researches with it's fascinating electronic-transport properties like quantum Hall Effect at room temperature, very high carrier mobility, long mean free path and ballistic transport of electrons.\cite{geim2007} In addition to this, superior mechanical strength, high thermal conductivity and remarkable flexibility of graphene makes it an ideal candidate for device applications.

The origin of exotic electronic-transport properties of graphene lies in it's linear energy dispersion (resembling the Dirac spectrum of massless fermions) at the high-symmetry points located at the six corners (denoted as K points) of the hexagonal Brillouin zone.\cite{wallace1947,rmp2009} Since the highest occupied and lowest unoccupied band touches each other at the Dirac points (K points), graphene is classified as a semi-metal. Unfortunately, lack of intrinsic bandgap limits the use of graphene to some extent. For example, the advantage of ultrahigh electron mobility is nullified by high off current in  graphene based field effect transistor (FET) devices. 

Fabrication of nanoribbons and quantum dots is one possible solution, as bandgap appears due to quantum confinement effect.\cite{son2006} Although graphene nanoribbons have several interesting features, like spontaneous spin polarization along the edges, their fabrication with atomically controlled edge shapes remains a challange.\cite{sonnat2006,bhowmick2010,bhowmick2013} Other alternative is to functionalize graphene via chemical adsorption of hydrogen\cite{sluiter2003,sofo2007} or halogen atoms like fluorine\cite{robinson2010,nair2010,radek2010,withers2010} and chlorine.\cite{li2011,garcia2011,sahin2012} Unfortunately, this leads to a large bandgap of magnitude 3.5-3.7 eV in case of hydrogenation, 2.9-3.1 eV in case of fluorination and 1.2-1.5 eV in case of chlorination, as reported in several computational and experimental studies.\cite{sluiter2003,sofo2007,robinson2010,nair2010,radek2010,withers2010,li2011,garcia2011,sahin2012} Reducing the bandgap is certainly going to make them more appealing for device applications.

Being atomically thin, 2D materials can not screen the external fields very effectively. As a result, perturbations like applied strain and electric field are known to significantly change the electronic band structure of 2D materials. For example, in case of MoS$_2$, experimental studies have reported nearly 100 meV bandgap decrease per percent of applied strain, accompanied by direct-to-indirect transition of the character of the bandgap.\cite{conley2013,lloyd2016} Similar predictions have been made for 2D phosphorus allotropes, where applied strain is also found to switch the preferred conduction direction.\cite{zhu2014,guan2014,fei2014} Bandgap can also be tuned by applying an electric field in a direction perpendicular to the plane of the 2D material, as shown for MoS$_2$ \cite{lupccp2014,qiapl2013,liujpcc2012}, phosphorene\cite{liunl2015,dai2014,dong2017,ghosh2015} and multilayer graphene.\cite{tang2011}

In this paper we select chlorographene, which has the lowest bandgap among the functionalized graphene siblings and study the effect of strain (within the elastic limit) on it's electonic band structure. We show that, sizeable bandgap of chlorographene can be reduced to a vanishingly small value of a few meV (under tensile strain) and it can even be converted to a metallic state (under compressive strain). We find that spin-orbit coupling is responsible for the few meV bandgap observed under high tensile strain, otherwise which appears like a gapless semi-metallic state of chlorographene. Based on symmetry arguments, we expect the results to be qualitatively true for other materials with same space group symmetry and in that sense, our study is quite general in nature.

\begin{figure*}[]
\centerline{\includegraphics[width=1.0 \columnwidth]{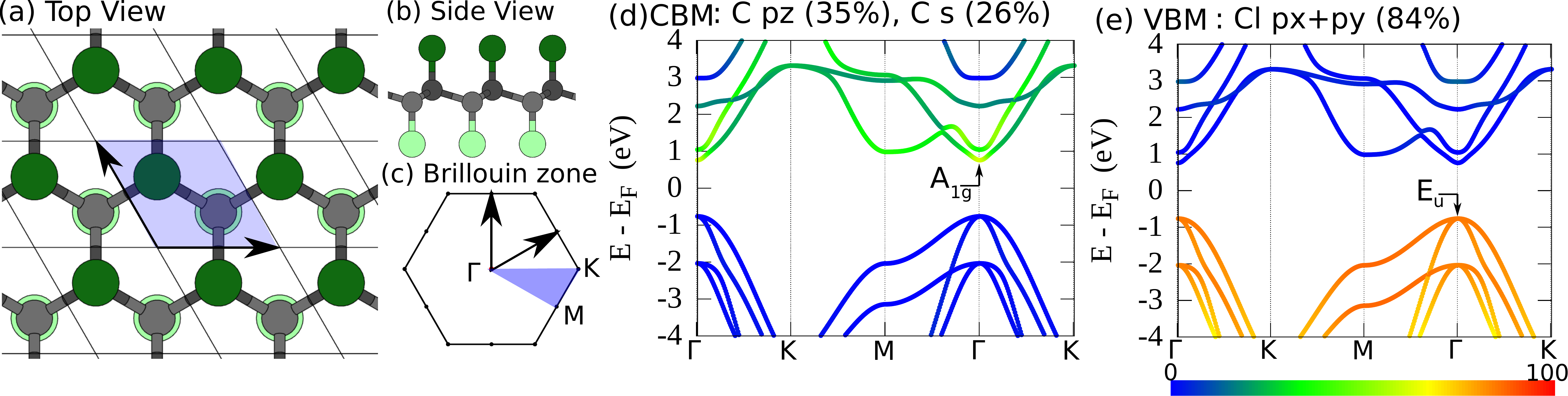}}
\caption{(a) Top view and (b) side view of chlorographene unit cell and (c) the corresponding high symmetry points in the reciprocal lattice. Carbon and chlorine atoms are shown in black and green, respectively, while the dark and light shading in panels (a) and (b) are done according to the height of the atoms. Crystal structures are prepared by using Xcrysden.\cite{Kokalj03,Kokalj99} Orbital resolved electronic band structure of chlorographene is depicted in (d) and (e), highlighting the major contributions from different atomic orbitals. Panel (d) illustrates the relative weight of C $p_z$ and C $s$ orbital on different energy bands. Panel (e) illustrates the same for Cl $p_x$ and $p_y$.  At the $\Gamma$ point, VBM (CBM) corresponds to the irreducible representation E$_u$ (A$_{1g}$) of D$_{3d}$ point group. VBM (E$_u$) is found to be doubly degenerate.}
\label{fig1}
\end{figure*}

Structural optimizations and electronic band structure calculations are carried out using QUANTUM ESPRESSO\cite{Gianozzi09} package, implementing density functional theory (DFT) using a plane-wave basis set (kinetic energy cutoff taken to be 80 Ry). Core electrons are treated using the norm-conserving pseudopotentials and exchange-correlation effects are included within the framework of generalized gradient approximations(GGA). A $k-$point mesh of $24\times 24\times 1$ is used to obtain the electron density in a self-consistent manner. Structural optimizations are carried out untill the energy difference between two succesive steps are less than 10$^{-4}$ Ry and all three components of the force on each atom are less than 10$^{-3}$ Ry/Bohr. 

Graphene has a honeycomb lattice of carbon atoms with a space group symmetry of $P6/mmm$ (\#191), which is lowered to $P\bar{3}m1$ (\#164) by full chlorination (one Cl atom per C) and the resulting 2D material is known as chlorographene. The crystal structure and hexagonal unit cell of chlorographene is illustrated in Fig~\ref{fig1}(a)-(b). Using the computational parameters mentioned above, the lattice constant of fully relaxed structure of chlorographene is measured to be 2.91 \AA, while the C-C and C-Cl bond length is found to be 1.75 \AA~ and 1.74 \AA, respectively and the C-C-C and C-C-Cl bond angle is obtained to be 111.98$^\circ$ and 106.83$^\circ$, respectively. The first Brillouin zone of chlorographene is shown in Fig~\ref{fig1}(c) and the high symmetry points are marked as $\Gamma$ (center of the Brillouin zone), K (corner of the Brillouin zone) and M (center of the edges of the Brillouin zone). 

The electronic band structure is plotted along the high symmetry lines $\Gamma$-K, K-M and M-$\Gamma$ [see Fig.~\ref{fig1}(d)-(e)].\footnote{Point group symmetry of the wave vector $\bm{k}$ at the $\Gamma$, K and  M point is D$_{3d}$, D$_3$ and C$_{2h}$, respectively and along the $\Gamma$-K, K-M and M-$\Gamma$ line is C$_2$, C$_2$ and C$_s$, respectively.} As shown in the diagram, a direct bandgap at the $\Gamma$ point, measuring 1.5 eV, is obtained in case of chlorographene,  which is in good agreement with the value reported in the literature.\cite{sahin2012} It is observed that the valence band maximum (VBM) is two fold degenerate (heavy hole and light hole), while the conduction band minimum (CBM) is non-degenerate. Degeneracy of the hole bands at the $\Gamma$ point is dictated by symmetry and it is also observed in other 2D materials of $P\bar{3}m1$ space group, like monolayer $\beta$ phosphorus\cite{zhu2014} and arsenic,\cite{Mardanya:2016} as well as blue phosphorus oxide.\cite{zhu2016} Two states at the VBM (one state at the CBM) corresponds to the irreducible representation E$_u$ (A$_{1g}$) of D$_{3d}$ point group at the $\Gamma$ point. When band states are projected on the atomic orbitals, it is found that, close to the $\Gamma$ point, the valence bands are mostly formed by the $p_x$ and $p_y$ orbital of Cl, while the conduction band edge is constituted mainly by the $p_z$ and $s$ orbitals of C.

\begin{figure*}[]
\centerline{\includegraphics[width=1.0 \columnwidth]{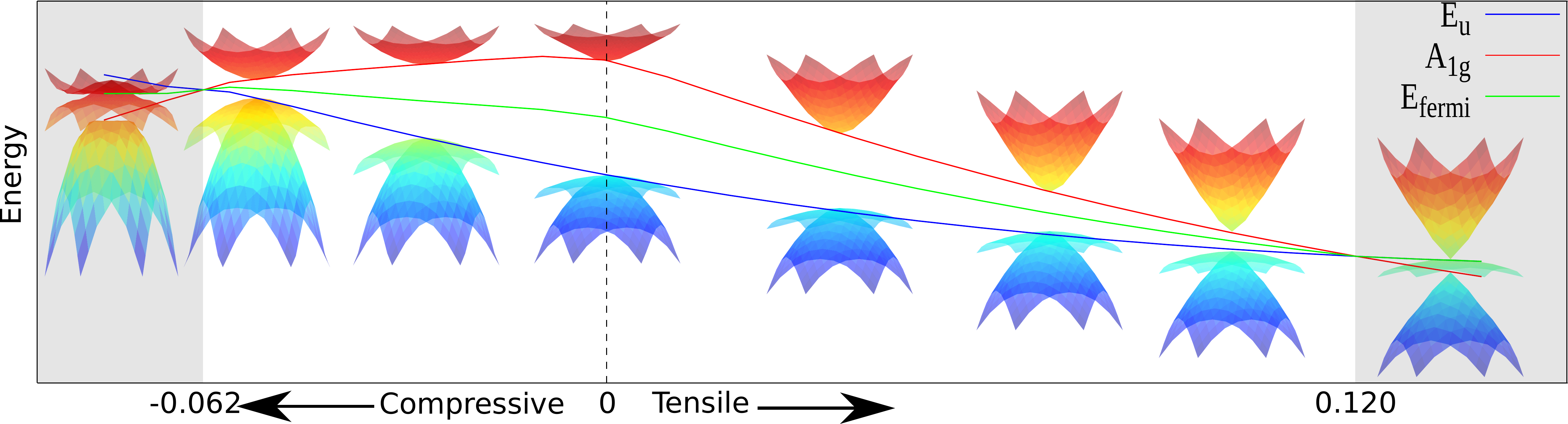}}
\caption{Effect of strain on band edges of chlorographene is shown. Energy level (measured with respect to the vacuum energy level) of the valence band edge (E$_u$ irreducible representation of D$_{3d}$ point group at the $\Gamma$ point) falls with increasing tensile strain, while it rises with increasing compressive strain. On the other hand, energy level of the conduction band edge (A$_{1g}$ irreducible representation of D$_{3d}$ point group at the $\Gamma$ point) falls with increasing tensile, as well as compressive strain. As a result of band edge shifting, bandgap of chlorographene closes gradually with increasing strain, followed by a phase transition (shaded region), which is qualitatively different for tensile and compressive strain. In the tensile side, a semiconductor to semimetal phase transition is observed beyond 12\% critical strain, while a semiconductor to metal phase transition is found beyond 6.2\% compressive strain.}
\label{fig2}
\end{figure*}

Impact of deformation on electronic band structure is investigated by applying compressive (upto 7\%) and tensile (upto 13\%) strain bi-axially, such that the symmetry of the pristine material is preserved after contraction or expansion of the unit cell vectors. The strains applied are within the elastic limit predicted for chlorographene, as predicted by \textit{ab initio} calculations.\cite{sahin2012} Bi-axial strain upto 6\% has so far been reported experimentally in similar 2D materials like MoS$_2$.\cite{lloyd2016} The overall effect of bi-axial strain is summarized in a three dimensional electronic band structure diagram, where only three low energy bands near the valence and conduction band edge of chlorographene are shown for the sake of clarity [see Fig.~\ref{fig2}]. In this diagram, we plot the energy level shift (measured with respect to the vacuum energy level) of E$_u$ and A$_{1g}$ states, as a function of applied strain. Clearly, energy level of doubly degenerate E$_u$ states falls with increasing tensile strain, while it rises with increasing compressive strain. In contrast, energy level of A$_{1g}$ states falls with increasing tensile, as well as compressive strain. Due to the energy level shift of E$_u$ and A$_{1g}$ states, bandgap of chlorographene gradually closes with increasing strain, which finally leads to a phase transition. Shaded areas, located extreme right and left of Fig.~\ref{fig2}, mark the regions where chlorographene is no longer a semiconductor. Interestingly, phase transition under tensile strain is qualitatively different from what we observe with compressive strain. It is found that, while a semiconductor to semi-metal transition takes place in the tensile region (beyond 12\% strain), on the other hand, chlorographene transforms to a metallic state under compression (beyond 6.2\% strain). Interestingly, two bands are also found to be touching each other along a circular line surrounding the $\Gamma$ point in the latter state.

\begin{figure}[]
\centerline{\includegraphics[width=1.0 \columnwidth]{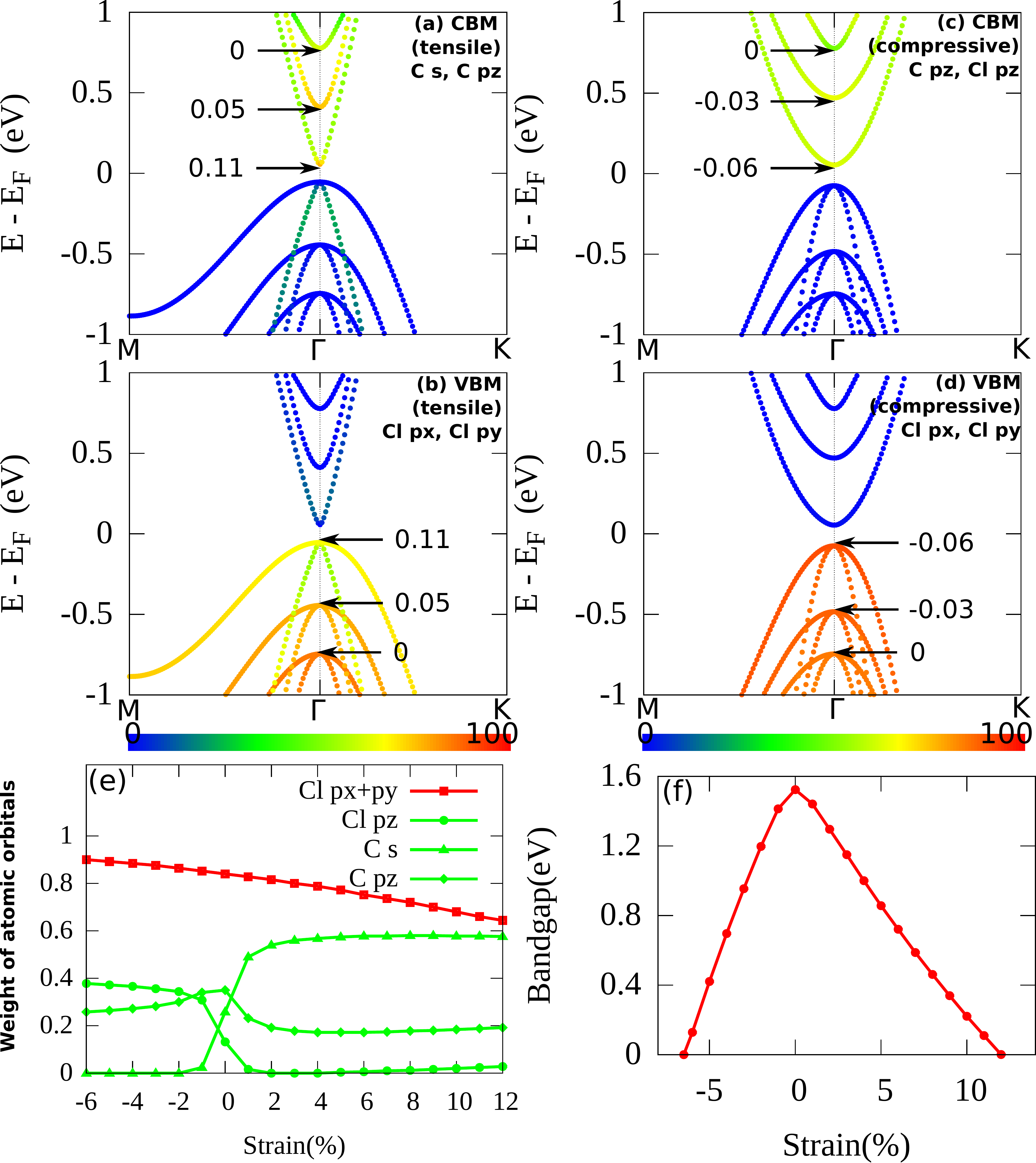}}
\caption{Orbital resolved band structure of chlorographene is illustrated; under (a)-(b) tensile (0, 5 and 11\%) and (c)-(d) compressive (0, 3 and 6\%) strain. The corresponding atomic orbitals are mentioned in respective panels. (e) Weight of the atomic orbitals, forming the band edges of chlorographene, is shown as a function of strain. (f) Bandgap of chlorographene is plotted at different values of strain.}
\label{fig3}
\end{figure}

Let us first have a detailed discussion on what happens until the point of phase transition. As long as chlorographene exists as a semiconductor, doubly degenerate E$_u$ states remain at the valence band top, while A$_{1g}$ state prevails at the conduction band bottom. The orbital resolved band structure of chlorographene under tensile strain is shown in Fig~\ref{fig3} (a) and (b). Clearly, near the $\Gamma$ point, the valence band edge is formed by mainly $p_x$ and $p_y$ atomic orbital of Cl. On the other hand, conduction band edge is composed of $s$ and $p_z$ orbital of C. In the same figure, the orbital resolved band structure of chlorographene under compressive strain is plotted in panel (c) and (d). A thorough comparison reveals that, the valence band edge is still dominated by the $p_x$ and $p_y$ atomic orbital of Cl, similar to the case of chlorographene under tensile strain. However, a significant contribution of $p_z$ orbital of Cl is found near the conduction band edge under compression; instead of $s$ orbital of C, as observed in case of tensile strain. Other than that, $p_z$ orbital of C has a significant weight in case of CBM of chlorographene, both under compression and tension. A summary of relevant atomic orbitals is given in Fig~\ref{fig3} (e) by plotting their weight as a function of strain. Clearly, in equilibrium (zero strain) weight of the atomic orbitals at the CBM is in the following order: C $p_z > $ C $s >$ Cl $p_z$. However, contribution from $s$ orbital of C atom rises and $p_z$ orbital of Cl atom falls rapidly under tensile strain. This is exactly opposite to what happens under compressive strain, where weight of $p_z$ orbital of Cl atom rises to prominence, while it decays to negligible values for $s$ orbital of C atom. The electronic band structures presented in Fig.~\ref{fig3} also clearly illustrates that the bandgap decreases with increasing strain (both compressive and tensile) and it's magnitude is plotted as a function of strain in panel (f) of the same figure. Note that, the bandgap changes more sharply under compressive strain, although ultimately it drops to zero in either case and the particular nature of the phase transition is going to be discussed in the following paragraphs. 

\begin{figure}[]
\centerline{\includegraphics[width=1.0 \columnwidth]{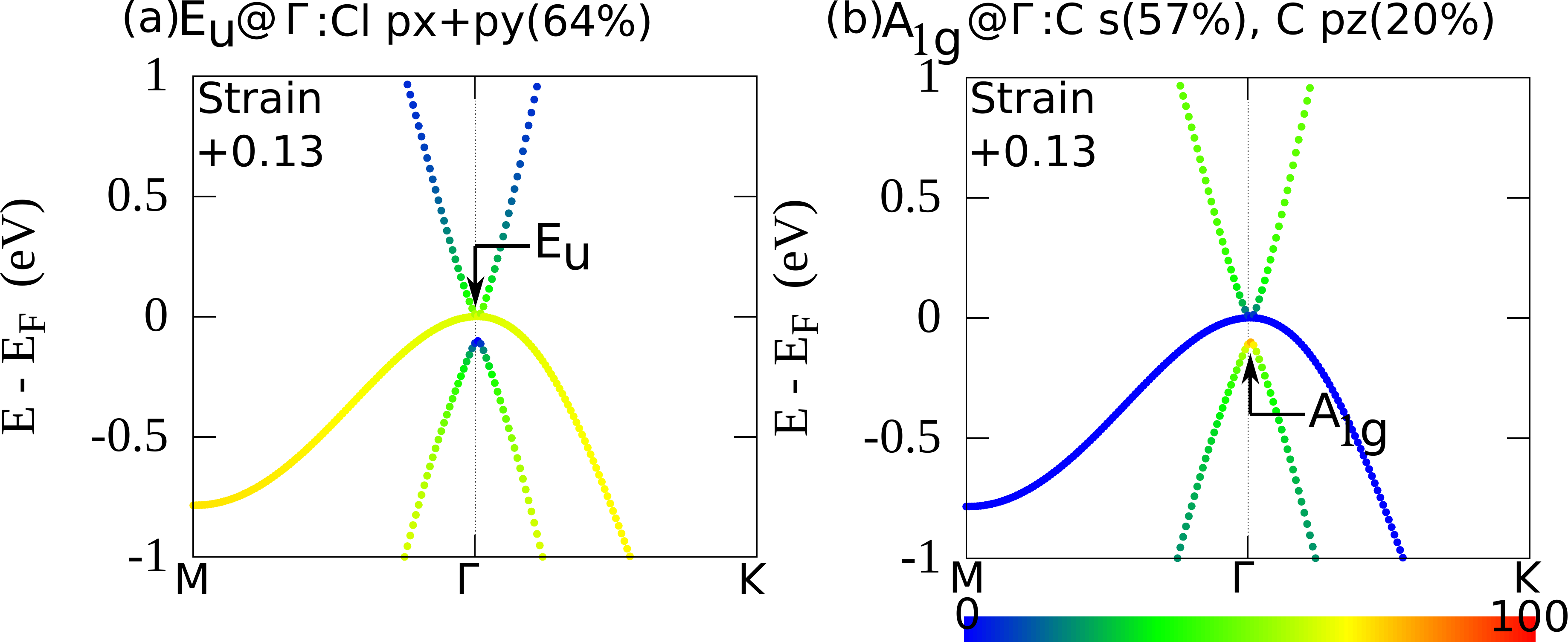}}
\caption{Orbital resolved band structure of chlorographene under $13\%$ tensile strain is shown in this figure. The valence and conduction band touches at the $\Gamma$ point (zero bandgap) and chlorographene exists in semi-metallic state.}
\label{fig4}
\end{figure}

First, let us focus on the semi-metallic phase appearing beyond 12\% tensile strain. A representative electronic band structure of semi-metallic phase of chlorographene (under 13\% tensile strain) is shown in Fig.~\ref{fig4}. Clearly, the valence band top and conduction band bottom is touching each other at the $\Gamma$ point. This originates from the two fold degeneracy of the E$_u$ states at the $\Gamma$ point. As discussed previously, E$_u$ states are doubly degenerate in equilibrium (zero strain) also. It is not surprising that the degeneracy survives, because uniform biaxial strain does not change the symmetry of chlorographene. As shown in Fig.~\ref{fig4}, energy level of E$_u$ is higher than that of A$_{1g}$ states, which is exactly opposite to what is observed in semiconducting state of chlorographene, i.e., in equilibrium (zero strain) and all the way to 12\% tensile strain. At this critical value of strain, A$_{1g}$ state drops below the energy level of E$_u$ states, which marks the onset of semi-metallic state of chlorographene [also see Fig.~\ref{fig2}]. Note that, in-between the semiconducting and semi-metallic phase, there might exist a triply degenerate state (two E$_u$ and one A$_{1g}$ state at the same energy level) at some particular value of strain, although it is difficult identify it because DFT can not accurately predict bandgap smaller than a few milli-electron volt. On the other hand, semi-metallic state beyond 12\% strain is very robust because it originates from the two fold degeneracy of the E$_u$ states, protected by the symmetry of chlorographene. Based on the orbital resolved band structure shown in Fig.~\ref{fig4}, it is also clear that the E$_u$ (A$_{1g}$) states at the $\Gamma$ point are mainly composed of $p_x$ and $p_y$ orbital of Cl ($s$ and $p_z$ orbital of C), which is consistent with the trend observed for chlorographene under tensile strain [see Fig.~\ref{fig3}(e)].

\begin{figure}[]
\centerline{\includegraphics[width=1.0 \columnwidth]{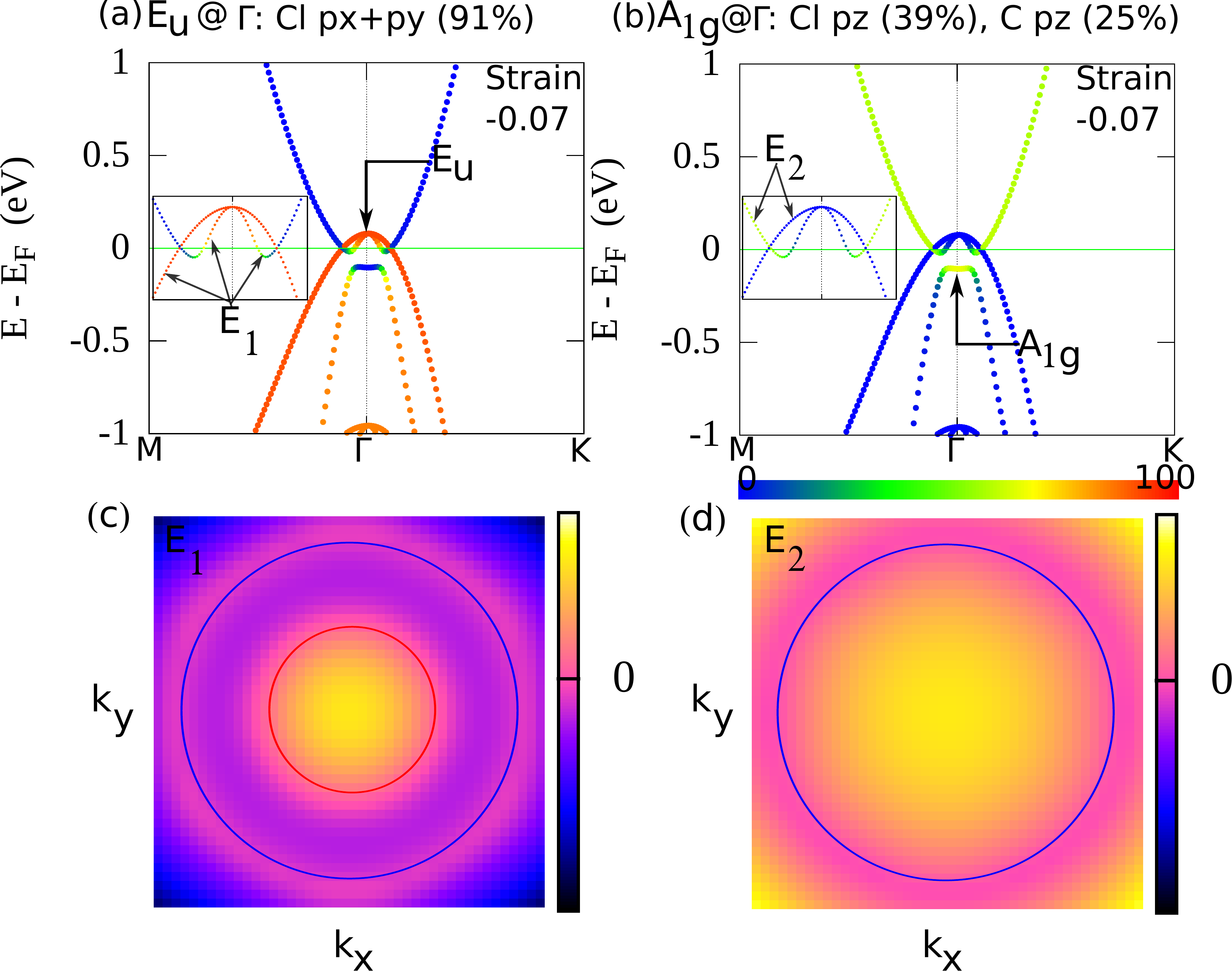}}
\caption{Orbital resolved band structure of metallic chlorographene (under $7\%$ compressive strain) is shown in (a) and (b). Relevant portion of the two bands (at the Fermi level) are magnified and shown in the insets. Band diagrams illustrated in the insets of (a) and (b) are further represented by surface plots of (c) E$_1(k_x,k_y)$ and (d) E$_2(k_x,k_y)$, where E$_2 \ge$ E$_1$ at a reciprocal lattice point $(k_x,k_y)$. Since E$_1$ and E$_2$ are calculated with respect to the E$_F$, Fermi level lies along the zero energy contour lines, illustrated by the blue and red circle. Two bands cross along the outer blue circle, as shown in panel (c) and (d). The inner red circle of panel (c) is the other zero energy contour line, related to the partially filled band of chlorographene.}
\label{fig5}
\end{figure}

Finally, we discuss the phase transition under compression and as mentioned previously, it is qualitatively different from the transition taking place under tensile strain [also see Fig.~\ref{fig2}]. Finer details are illustrated in Fig.~\ref{fig5} (a) and (b), where electronic bands of chlorographene under 7\% compressive strain are plotted along various high symmetry lines. Again, since uniform biaxial strain preserves the symmeties of chlorographene crystal, the doubly degenerate E$_u$ states at the $\Gamma$ point persists and have higher energy than that of A$_{1g}$ states. Thus, similar to tensile strain, phase transition under compression also happens as the energy of A$_{1g}$ drops below that of E$_u$ states [see Fig.~\ref{fig2}]. Orbital resolved band structure shown in Fig.~\ref{fig5} (a) and (b) also reveals that the E$_u$ (A$_{1g}$) states at the $\Gamma$ point are mainly composed of $p_x$ and $p_y$ orbital of Cl ($p_z$ orbital of Cl and $p_z$ orbital of C), which is consistent with the trend observed for chlorographene under compressive strain [see Fig.~\ref{fig3}(e)]. Note that, neither E$_u$, nor A$_{1g}$ states at the $\Gamma$ point are at the Fermi level, which lies in between the two [see Fig.~\ref{fig5}(a) and (b)]. This leads to a partially filled band and consequently chlorographene transforms from semiconducting to a metallic state. Note that, Fremi level crosses the energy bands in two distinct places between $\Gamma$M, as well as $\Gamma$K [see insets of Fig.~\ref{fig5} (a) and (b)]. This is further elucidated by using surface plots of E$_1(k_x,k_y)$ and E$_2(k_x,k_y)$, where E$_2 \ge$ E$_1$ at a reciprocal lattice point $(k_x,k_y)$ [see Fig.~\ref{fig5}(c)-(d)]. Fermi energy (zero energy contour line) lies along the red and blue circle, as shown in Fig.~\ref{fig5}(c) and (d). Among the two, the outer circle (blue) has very interesting feature, with two energy bands crossing each other along the line. The inner red circle in Fig.~\ref{fig5}(c) originates from the partially filled band [see Fig.~\ref{fig5}(a)] and shows the location of the points where Fermi energy crosses E$_1(k_x,k_y)$.

\begin{figure*}[]
\centerline{\includegraphics[width=1.0 \columnwidth]{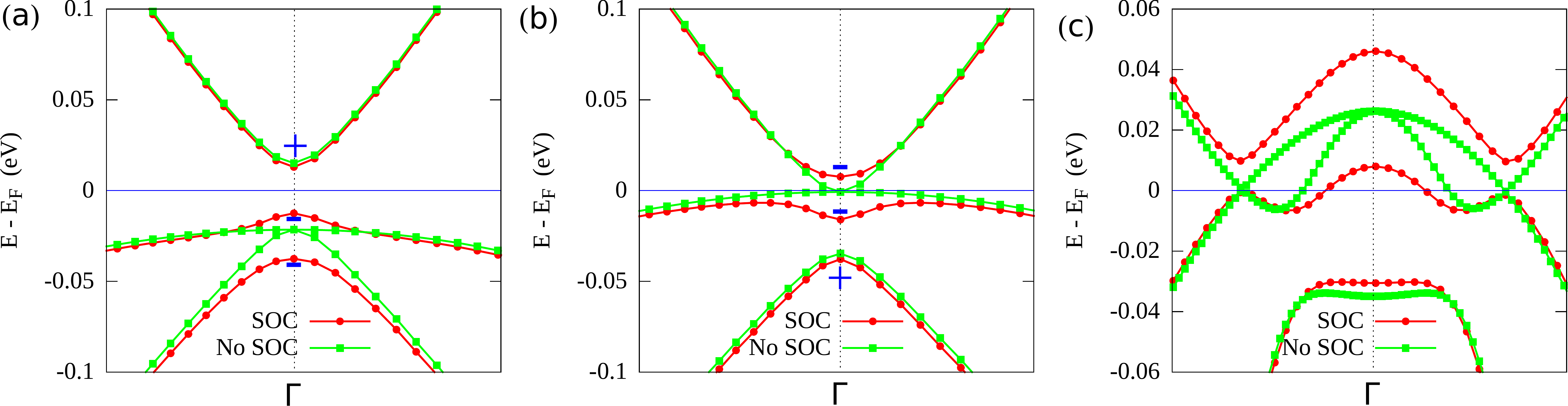}}
\caption{Effect of spin-orbit coupling (SOC) on electronic band structure of chlorographene is shown for (a) 11\% tensile, (b) 13\% tensile and (c) 7\% compressive strain. Clearly, degeneracy of the E$_u$ states at the $\Gamma$ point is lifted and a gap of magnitude 20 and 40 meV is created, in case of tensile and compressive strain, respectively. Additionally, in case of chlorographene under 7\% compressive strain, a 10 meV energy gap is opened along the outer blue circle, where two bands were touching each other [see Fig.~\ref{fig5}]. In panel (a) and (b), parities of the Bloch states at the $\Gamma$ point are labeled by + and -.}
\label{fig6}
\end{figure*}

So far our calculations do not include the effect of spin-orbit coupling. This is known to lift the degeneracy and induce energy gaps, especially at the high-symmetry points, in other 2D material like graphene.\cite{sergej2010} As discussed previously, the two $p$ orbitals of Cl, which makes the VBM of chlorographene (when it is a semiconductor), are degenerate at the $\Gamma$  point [see Fig~\ref{fig1} and Fig~\ref{fig3}]. However, once the spin-orbit coupling\footnote{Fully relativistic pseudopotentials are used for spin-orbit coupling calculations.} is turned on, the degeneracy of the  E$_u$ states is lifted at the $\Gamma$ point and the energy levels are split by a small margin of 20--40 meV. This is observed irrespective of the magnitude of the applied strain, including the equilibrium (i.e., zero strain) state, as all of them have same space group symmetry. Three different cases are illustrated in Fig.~\ref{fig6}: (a) 11\% and (b) 13\% stretched, which is before and after transition under tensile strain and (c) 7\% squeezed, which is beyond the transition point under compressive strain. Interestingly, what appeared as a semi-metallic state of chlorographene beyond 12\% tensile strain; after inclusion of spin-orbit coupling, turns out to be a semiconducting phase with a vanishingly small bandgap, measuring $\sim$ 20 meV [see Fig.~\ref{fig6} (b)]. Further, comparing panel (a) and (b) of Fig.~\ref{fig6}, a parity exchange between the occupied and unoccupied bands is observed at the $\Gamma$ point. This type of parity change is known to be associated with topological phase transition.\cite{Fu2007} The metallic phase of chlorographene, which appears beyond 6.2\% compressive strain, remains as it is even with spin-orbit coupling, because of the partially filled band [see Fig.~\ref{fig6}(c)]. However, a small energy gap ($\sim$ 10 meV) opens along the outer blue circle [see Fig.~\ref{fig5} (c) and (d) and compare with Fig.~\ref{fig6}(c)] due to the spin-orbit coupling.

Note that, bandgap values reported in this work are obtained from DFT-GGA calculations, which is known to underestimate it's magnitude. As reported in the literature, using rigorous GW many-body perturbation theory method, bandgap of chlorographene (DFT-GGA estimate equal to 1.5 eV) was found to increase dramatically to a range of 4.3 to 4.9 eV.\cite{jctc2013,sahin2012} However, GW correction generally overestimates the bandgap.\cite{crowley2016} For example, experimental bandgap of fluorographene lies in the range of 3 to 3.8 eV, but GW calculations predict it to be around 7 to 8 eV, while the DFT-GGA predictions are very close to the experimental value.\cite{jctc2013} However, even a 50-60\% increase of bandgap of chlorographene is only going to enhance the strain at which the transformations are observed, while the results reported in this paper remains qualitatively same.

In conclusion, we show that chlorographene, which has a sizeable bandgap, can be driven to a state of vanishingly small bandgap or even a metal, by applying external strain. The semiconductor to metal transition takes place under compression, while tensile strain reduces the bandgap to a vanishingly small value of a few meV. Two bands are also found to be touching each other (at certain points lying on a circle surrounding the $\Gamma$ point in the Brillouin zone) under sufficiently large compression, although they split marginally due to spin-orbit coupling.

\section{Acknowledgements}
SB acknowledges funding from SERB Fast Track Scheme for Young Scientist (SB/FTP/ETA-0036/2014). We also thank CC IITK for providing HPC facility. 

\bibliography{ref}

\begin{thebibliography}{43}%
\makeatletter
\providecommand \@ifxundefined [1]{%
 \@ifx{#1\undefined}
}%
\providecommand \@ifnum [1]{%
 \ifnum #1\expandafter \@firstoftwo
 \else \expandafter \@secondoftwo
 \fi
}%
\providecommand \@ifx [1]{%
 \ifx #1\expandafter \@firstoftwo
 \else \expandafter \@secondoftwo
 \fi
}%
\providecommand \natexlab [1]{#1}%
\providecommand \enquote  [1]{``#1''}%
\providecommand \bibnamefont  [1]{#1}%
\providecommand \bibfnamefont [1]{#1}%
\providecommand \citenamefont [1]{#1}%
\providecommand \href@noop [0]{\@secondoftwo}%
\providecommand \href [0]{\begingroup \@sanitize@url \@href}%
\providecommand \@href[1]{\@@startlink{#1}\@@href}%
\providecommand \@@href[1]{\endgroup#1\@@endlink}%
\providecommand \@sanitize@url [0]{\catcode `\\12\catcode `\$12\catcode
  `\&12\catcode `\#12\catcode `\^12\catcode `\_12\catcode `\%12\relax}%
\providecommand \@@startlink[1]{}%
\providecommand \@@endlink[0]{}%
\providecommand \url  [0]{\begingroup\@sanitize@url \@url }%
\providecommand \@url [1]{\endgroup\@href {#1}{\urlprefix }}%
\providecommand \urlprefix  [0]{URL }%
\providecommand \Eprint [0]{\href }%
\providecommand \doibase [0]{http://dx.doi.org/}%
\providecommand \selectlanguage [0]{\@gobble}%
\providecommand \bibinfo  [0]{\@secondoftwo}%
\providecommand \bibfield  [0]{\@secondoftwo}%
\providecommand \translation [1]{[#1]}%
\providecommand \BibitemOpen [0]{}%
\providecommand \bibitemStop [0]{}%
\providecommand \bibitemNoStop [0]{.\EOS\space}%
\providecommand \EOS [0]{\spacefactor3000\relax}%
\providecommand \BibitemShut  [1]{\csname bibitem#1\endcsname}%
\let\auto@bib@innerbib\@empty
\bibitem [{\citenamefont {Bhimanapati}\ \emph {et~al.}(2015)\citenamefont
  {Bhimanapati}, \citenamefont {Lin}, \citenamefont {Meunier}, \citenamefont
  {Jung}, \citenamefont {Cha}, \citenamefont {Das}, \citenamefont {Xiao},
  \citenamefont {Son}, \citenamefont {Strano}, \citenamefont {Cooper},
  \citenamefont {Liang}, \citenamefont {Louie}, \citenamefont {Ringe},
  \citenamefont {Zhou}, \citenamefont {Kim}, \citenamefont {Naik},
  \citenamefont {Sumpter}, \citenamefont {Terrones}, \citenamefont {Xia},
  \citenamefont {Wang}, \citenamefont {Zhu}, \citenamefont {Akinwande},
  \citenamefont {Alem}, \citenamefont {Schuller}, \citenamefont {Schaak},
  \citenamefont {Terrones},\ and\ \citenamefont {Robinson}}]{ganesh2015}%
  \BibitemOpen
  \bibfield  {author} {\bibinfo {author} {\bibfnamefont {Ganesh~R.}\
  \bibnamefont {Bhimanapati}}, \bibinfo {author} {\bibfnamefont {Zhong}\
  \bibnamefont {Lin}}, \bibinfo {author} {\bibfnamefont {Vincent}\ \bibnamefont
  {Meunier}}, \bibinfo {author} {\bibfnamefont {Yeonwoong}\ \bibnamefont
  {Jung}}, \bibinfo {author} {\bibfnamefont {Judy}\ \bibnamefont {Cha}},
  \bibinfo {author} {\bibfnamefont {Saptarshi}\ \bibnamefont {Das}}, \bibinfo
  {author} {\bibfnamefont {Di}~\bibnamefont {Xiao}}, \bibinfo {author}
  {\bibfnamefont {Youngwoo}\ \bibnamefont {Son}}, \bibinfo {author}
  {\bibfnamefont {Michael~S.}\ \bibnamefont {Strano}}, \bibinfo {author}
  {\bibfnamefont {Valentino~R.}\ \bibnamefont {Cooper}}, \bibinfo {author}
  {\bibfnamefont {Liangbo}\ \bibnamefont {Liang}}, \bibinfo {author}
  {\bibfnamefont {Steven~G.}\ \bibnamefont {Louie}}, \bibinfo {author}
  {\bibfnamefont {Emilie}\ \bibnamefont {Ringe}}, \bibinfo {author}
  {\bibfnamefont {Wu}~\bibnamefont {Zhou}}, \bibinfo {author} {\bibfnamefont
  {Steve~S.}\ \bibnamefont {Kim}}, \bibinfo {author} {\bibfnamefont
  {Rajesh~R.}\ \bibnamefont {Naik}}, \bibinfo {author} {\bibfnamefont
  {Bobby~G.}\ \bibnamefont {Sumpter}}, \bibinfo {author} {\bibfnamefont
  {Humberto}\ \bibnamefont {Terrones}}, \bibinfo {author} {\bibfnamefont
  {Fengnian}\ \bibnamefont {Xia}}, \bibinfo {author} {\bibfnamefont {Yeliang}\
  \bibnamefont {Wang}}, \bibinfo {author} {\bibfnamefont {Jun}\ \bibnamefont
  {Zhu}}, \bibinfo {author} {\bibfnamefont {Deji}\ \bibnamefont {Akinwande}},
  \bibinfo {author} {\bibfnamefont {Nasim}\ \bibnamefont {Alem}}, \bibinfo
  {author} {\bibfnamefont {Jon~A.}\ \bibnamefont {Schuller}}, \bibinfo {author}
  {\bibfnamefont {Raymond~E.}\ \bibnamefont {Schaak}}, \bibinfo {author}
  {\bibfnamefont {Mauricio}\ \bibnamefont {Terrones}}, \ and\ \bibinfo {author}
  {\bibfnamefont {Joshua~A.}\ \bibnamefont {Robinson}},\ }\bibfield  {title}
  {\enquote {\bibinfo {title} {Recent advances in two-dimensional materials
  beyond graphene},}\ }\href {\doibase 10.1021/acsnano.5b05556} {\bibfield
  {journal} {\bibinfo  {journal} {ACS Nano}\ }\textbf {\bibinfo {volume} {9}},\
  \bibinfo {pages} {11509--11539} (\bibinfo {year} {2015})}\BibitemShut
  {NoStop}%
\bibitem [{\citenamefont {Novoselov}\ \emph {et~al.}(2004)\citenamefont
  {Novoselov}, \citenamefont {Geim}, \citenamefont {Morozov}, \citenamefont
  {Jiang}, \citenamefont {Zhang}, \citenamefont {Dubonos}, \citenamefont
  {Grigorieva},\ and\ \citenamefont {Firsov}}]{Novoselov666}%
  \BibitemOpen
  \bibfield  {author} {\bibinfo {author} {\bibfnamefont {K.~S.}\ \bibnamefont
  {Novoselov}}, \bibinfo {author} {\bibfnamefont {A.~K.}\ \bibnamefont {Geim}},
  \bibinfo {author} {\bibfnamefont {S.~V.}\ \bibnamefont {Morozov}}, \bibinfo
  {author} {\bibfnamefont {D.}~\bibnamefont {Jiang}}, \bibinfo {author}
  {\bibfnamefont {Y.}~\bibnamefont {Zhang}}, \bibinfo {author} {\bibfnamefont
  {S.~V.}\ \bibnamefont {Dubonos}}, \bibinfo {author} {\bibfnamefont {I.~V.}\
  \bibnamefont {Grigorieva}}, \ and\ \bibinfo {author} {\bibfnamefont {A.~A.}\
  \bibnamefont {Firsov}},\ }\bibfield  {title} {\enquote {\bibinfo {title}
  {Electric field effect in atomically thin carbon films},}\ }\href {\doibase
  10.1126/science.1102896} {\bibfield  {journal} {\bibinfo  {journal}
  {Science}\ }\textbf {\bibinfo {volume} {306}},\ \bibinfo {pages} {666--669}
  (\bibinfo {year} {2004})}\BibitemShut {NoStop}%
\bibitem [{\citenamefont {Novoselov}\ \emph {et~al.}(2005)\citenamefont
  {Novoselov}, \citenamefont {Jiang}, \citenamefont {Schedin}, \citenamefont
  {Booth}, \citenamefont {Khotkevich}, \citenamefont {Morozov},\ and\
  \citenamefont {Geim}}]{Novoselov26072005}%
  \BibitemOpen
  \bibfield  {author} {\bibinfo {author} {\bibfnamefont {K.~S.}\ \bibnamefont
  {Novoselov}}, \bibinfo {author} {\bibfnamefont {D.}~\bibnamefont {Jiang}},
  \bibinfo {author} {\bibfnamefont {F.}~\bibnamefont {Schedin}}, \bibinfo
  {author} {\bibfnamefont {T.~J.}\ \bibnamefont {Booth}}, \bibinfo {author}
  {\bibfnamefont {V.~V.}\ \bibnamefont {Khotkevich}}, \bibinfo {author}
  {\bibfnamefont {S.~V.}\ \bibnamefont {Morozov}}, \ and\ \bibinfo {author}
  {\bibfnamefont {A.~K.}\ \bibnamefont {Geim}},\ }\bibfield  {title} {\enquote
  {\bibinfo {title} {Two-dimensional atomic crystals},}\ }\href {\doibase
  10.1073/pnas.0502848102} {\bibfield  {journal} {\bibinfo  {journal}
  {Proceedings of the National Academy of Sciences of the United States of
  America}\ }\textbf {\bibinfo {volume} {102}},\ \bibinfo {pages}
  {10451--10453} (\bibinfo {year} {2005})}\BibitemShut {NoStop}%
\bibitem [{\citenamefont {Geim}\ and\ \citenamefont
  {Novoselov}(2007)}]{geim2007}%
  \BibitemOpen
  \bibfield  {author} {\bibinfo {author} {\bibfnamefont {Andre~K}\ \bibnamefont
  {Geim}}\ and\ \bibinfo {author} {\bibfnamefont {Konstantin~S}\ \bibnamefont
  {Novoselov}},\ }\bibfield  {title} {\enquote {\bibinfo {title} {The rise of
  graphene},}\ }\href {\doibase 10.1038/nmat1849} {\bibfield  {journal}
  {\bibinfo  {journal} {Nat. Mater.}\ }\textbf {\bibinfo {volume} {6}},\
  \bibinfo {pages} {183--191} (\bibinfo {year} {2007})}\BibitemShut {NoStop}%
\bibitem [{\citenamefont {Wallace}(1947)}]{wallace1947}%
  \BibitemOpen
  \bibfield  {author} {\bibinfo {author} {\bibfnamefont {P.~R.}\ \bibnamefont
  {Wallace}},\ }\bibfield  {title} {\enquote {\bibinfo {title} {The band theory
  of graphite},}\ }\href {\doibase 10.1103/PhysRev.71.622} {\bibfield
  {journal} {\bibinfo  {journal} {Phys. Rev.}\ }\textbf {\bibinfo {volume}
  {71}},\ \bibinfo {pages} {622--634} (\bibinfo {year} {1947})}\BibitemShut
  {NoStop}%
\bibitem [{\citenamefont {Castro~Neto}\ \emph {et~al.}(2009)\citenamefont
  {Castro~Neto}, \citenamefont {Guinea}, \citenamefont {Peres}, \citenamefont
  {Novoselov},\ and\ \citenamefont {Geim}}]{rmp2009}%
  \BibitemOpen
  \bibfield  {author} {\bibinfo {author} {\bibfnamefont {A.~H.}\ \bibnamefont
  {Castro~Neto}}, \bibinfo {author} {\bibfnamefont {F.}~\bibnamefont {Guinea}},
  \bibinfo {author} {\bibfnamefont {N.~M.~R.}\ \bibnamefont {Peres}}, \bibinfo
  {author} {\bibfnamefont {K.~S.}\ \bibnamefont {Novoselov}}, \ and\ \bibinfo
  {author} {\bibfnamefont {A.~K.}\ \bibnamefont {Geim}},\ }\bibfield  {title}
  {\enquote {\bibinfo {title} {The electronic properties of graphene},}\ }\href
  {\doibase 10.1103/RevModPhys.81.109} {\bibfield  {journal} {\bibinfo
  {journal} {Rev. Mod. Phys.}\ }\textbf {\bibinfo {volume} {81}},\ \bibinfo
  {pages} {109--162} (\bibinfo {year} {2009})}\BibitemShut {NoStop}%
\bibitem [{\citenamefont {Son}\ \emph {et~al.}(2006)\citenamefont {Son},
  \citenamefont {Cohen},\ and\ \citenamefont {Louie}}]{son2006}%
  \BibitemOpen
  \bibfield  {author} {\bibinfo {author} {\bibfnamefont {Young-Woo}\
  \bibnamefont {Son}}, \bibinfo {author} {\bibfnamefont {Marvin~L.}\
  \bibnamefont {Cohen}}, \ and\ \bibinfo {author} {\bibfnamefont {Steven~G.}\
  \bibnamefont {Louie}},\ }\bibfield  {title} {\enquote {\bibinfo {title}
  {Energy gaps in graphene nanoribbons},}\ }\href {\doibase
  10.1103/PhysRevLett.97.216803} {\bibfield  {journal} {\bibinfo  {journal}
  {Phys. Rev. Lett.}\ }\textbf {\bibinfo {volume} {97}},\ \bibinfo {pages}
  {216803} (\bibinfo {year} {2006})}\BibitemShut {NoStop}%
\bibitem [{\citenamefont {{Son}}\ \emph {et~al.}(2006)\citenamefont {{Son}},
  \citenamefont {{Cohen}},\ and\ \citenamefont {{Louie}}}]{sonnat2006}%
  \BibitemOpen
  \bibfield  {author} {\bibinfo {author} {\bibfnamefont {Y.-W.}\ \bibnamefont
  {{Son}}}, \bibinfo {author} {\bibfnamefont {M.~L.}\ \bibnamefont {{Cohen}}},
  \ and\ \bibinfo {author} {\bibfnamefont {S.~G.}\ \bibnamefont {{Louie}}},\
  }\bibfield  {title} {\enquote {\bibinfo {title} {{Half-metallic graphene
  nanoribbons}},}\ }\href {\doibase 10.1038/nature05180} {\bibfield  {journal}
  {\bibinfo  {journal} {\nat}\ }\textbf {\bibinfo {volume} {444}},\ \bibinfo
  {pages} {347--349} (\bibinfo {year} {2006})},\ \Eprint
  {http://arxiv.org/abs/cond-mat/0611600} {cond-mat/0611600} \BibitemShut
  {NoStop}%
\bibitem [{\citenamefont {Bhowmick}\ and\ \citenamefont
  {Shenoy}(2010)}]{bhowmick2010}%
  \BibitemOpen
  \bibfield  {author} {\bibinfo {author} {\bibfnamefont {Somnath}\ \bibnamefont
  {Bhowmick}}\ and\ \bibinfo {author} {\bibfnamefont {Vijay~B.}\ \bibnamefont
  {Shenoy}},\ }\bibfield  {title} {\enquote {\bibinfo {title} {Weber-fechner
  type nonlinear behavior in zigzag edge graphene nanoribbons},}\ }\href
  {\doibase 10.1103/PhysRevB.82.155448} {\bibfield  {journal} {\bibinfo
  {journal} {Phys. Rev. B}\ }\textbf {\bibinfo {volume} {82}},\ \bibinfo
  {pages} {155448} (\bibinfo {year} {2010})}\BibitemShut {NoStop}%
\bibitem [{\citenamefont {Bhowmick}\ \emph {et~al.}(2013)\citenamefont
  {Bhowmick}, \citenamefont {Medhi},\ and\ \citenamefont
  {Shenoy}}]{bhowmick2013}%
  \BibitemOpen
  \bibfield  {author} {\bibinfo {author} {\bibfnamefont {Somnath}\ \bibnamefont
  {Bhowmick}}, \bibinfo {author} {\bibfnamefont {Amal}\ \bibnamefont {Medhi}},
  \ and\ \bibinfo {author} {\bibfnamefont {Vijay~B.}\ \bibnamefont {Shenoy}},\
  }\bibfield  {title} {\enquote {\bibinfo {title} {Sensory-organ-like response
  determines the magnetism of zigzag-edged honeycomb nanoribbons},}\ }\href
  {\doibase 10.1103/PhysRevB.87.085412} {\bibfield  {journal} {\bibinfo
  {journal} {Phys. Rev. B}\ }\textbf {\bibinfo {volume} {87}},\ \bibinfo
  {pages} {085412} (\bibinfo {year} {2013})}\BibitemShut {NoStop}%
\bibitem [{\citenamefont {Sluiter}\ and\ \citenamefont
  {Kawazoe}(2003)}]{sluiter2003}%
  \BibitemOpen
  \bibfield  {author} {\bibinfo {author} {\bibfnamefont {Marcel H.~F.}\
  \bibnamefont {Sluiter}}\ and\ \bibinfo {author} {\bibfnamefont {Yoshiyuki}\
  \bibnamefont {Kawazoe}},\ }\bibfield  {title} {\enquote {\bibinfo {title}
  {Cluster expansion method for adsorption: Application to hydrogen
  chemisorption on graphene},}\ }\href {\doibase 10.1103/PhysRevB.68.085410}
  {\bibfield  {journal} {\bibinfo  {journal} {Phys. Rev. B}\ }\textbf {\bibinfo
  {volume} {68}},\ \bibinfo {pages} {085410} (\bibinfo {year}
  {2003})}\BibitemShut {NoStop}%
\bibitem [{\citenamefont {Sofo}\ \emph {et~al.}(2007)\citenamefont {Sofo},
  \citenamefont {Chaudhari},\ and\ \citenamefont {Barber}}]{sofo2007}%
  \BibitemOpen
  \bibfield  {author} {\bibinfo {author} {\bibfnamefont {Jorge~O.}\
  \bibnamefont {Sofo}}, \bibinfo {author} {\bibfnamefont {Ajay~S.}\
  \bibnamefont {Chaudhari}}, \ and\ \bibinfo {author} {\bibfnamefont {Greg~D.}\
  \bibnamefont {Barber}},\ }\bibfield  {title} {\enquote {\bibinfo {title}
  {Graphane: A two-dimensional hydrocarbon},}\ }\href {\doibase
  10.1103/PhysRevB.75.153401} {\bibfield  {journal} {\bibinfo  {journal} {Phys.
  Rev. B}\ }\textbf {\bibinfo {volume} {75}},\ \bibinfo {pages} {153401}
  (\bibinfo {year} {2007})}\BibitemShut {NoStop}%
\bibitem [{\citenamefont {Robinson}\ \emph {et~al.}(2010)\citenamefont
  {Robinson}, \citenamefont {Burgess}, \citenamefont {Junkermeier},
  \citenamefont {Badescu}, \citenamefont {Reinecke}, \citenamefont {Perkins},
  \citenamefont {Zalalutdniov}, \citenamefont {Baldwin}, \citenamefont
  {Culbertson}, \citenamefont {Sheehan},\ and\ \citenamefont
  {Snow}}]{robinson2010}%
  \BibitemOpen
  \bibfield  {author} {\bibinfo {author} {\bibfnamefont {Jeremy~T.}\
  \bibnamefont {Robinson}}, \bibinfo {author} {\bibfnamefont {James~S.}\
  \bibnamefont {Burgess}}, \bibinfo {author} {\bibfnamefont {Chad~E.}\
  \bibnamefont {Junkermeier}}, \bibinfo {author} {\bibfnamefont {Stefan~C.}\
  \bibnamefont {Badescu}}, \bibinfo {author} {\bibfnamefont {Thomas~L.}\
  \bibnamefont {Reinecke}}, \bibinfo {author} {\bibfnamefont {F.~Keith}\
  \bibnamefont {Perkins}}, \bibinfo {author} {\bibfnamefont {Maxim~K.}\
  \bibnamefont {Zalalutdniov}}, \bibinfo {author} {\bibfnamefont {Jeffrey~W.}\
  \bibnamefont {Baldwin}}, \bibinfo {author} {\bibfnamefont {James~C.}\
  \bibnamefont {Culbertson}}, \bibinfo {author} {\bibfnamefont {Paul~E.}\
  \bibnamefont {Sheehan}}, \ and\ \bibinfo {author} {\bibfnamefont {Eric~S.}\
  \bibnamefont {Snow}},\ }\bibfield  {title} {\enquote {\bibinfo {title}
  {Properties of fluorinated graphene films},}\ }\href {\doibase
  10.1021/nl101437p} {\bibfield  {journal} {\bibinfo  {journal} {Nano Letters}\
  }\textbf {\bibinfo {volume} {10}},\ \bibinfo {pages} {3001--3005} (\bibinfo
  {year} {2010})}\BibitemShut {NoStop}%
\bibitem [{\citenamefont {Nair}\ \emph {et~al.}(2010)\citenamefont {Nair},
  \citenamefont {Ren}, \citenamefont {Jalil}, \citenamefont {Riaz},
  \citenamefont {Kravets}, \citenamefont {Britnell}, \citenamefont {Blake},
  \citenamefont {Schedin}, \citenamefont {Mayorov}, \citenamefont {Yuan},
  \citenamefont {Katsnelson}, \citenamefont {Cheng}, \citenamefont
  {Strupinski}, \citenamefont {Bulusheva}, \citenamefont {Okotrub},
  \citenamefont {Grigorieva}, \citenamefont {Grigorenko}, \citenamefont
  {Novoselov},\ and\ \citenamefont {Geim}}]{nair2010}%
  \BibitemOpen
  \bibfield  {author} {\bibinfo {author} {\bibfnamefont {Rahul~R.}\
  \bibnamefont {Nair}}, \bibinfo {author} {\bibfnamefont {Wencai}\ \bibnamefont
  {Ren}}, \bibinfo {author} {\bibfnamefont {Rashid}\ \bibnamefont {Jalil}},
  \bibinfo {author} {\bibfnamefont {Ibtsam}\ \bibnamefont {Riaz}}, \bibinfo
  {author} {\bibfnamefont {Vasyl~G.}\ \bibnamefont {Kravets}}, \bibinfo
  {author} {\bibfnamefont {Liam}\ \bibnamefont {Britnell}}, \bibinfo {author}
  {\bibfnamefont {Peter}\ \bibnamefont {Blake}}, \bibinfo {author}
  {\bibfnamefont {Fredrik}\ \bibnamefont {Schedin}}, \bibinfo {author}
  {\bibfnamefont {Alexander~S.}\ \bibnamefont {Mayorov}}, \bibinfo {author}
  {\bibfnamefont {Shengjun}\ \bibnamefont {Yuan}}, \bibinfo {author}
  {\bibfnamefont {Mikhail~I.}\ \bibnamefont {Katsnelson}}, \bibinfo {author}
  {\bibfnamefont {Hui-Ming}\ \bibnamefont {Cheng}}, \bibinfo {author}
  {\bibfnamefont {Wlodek}\ \bibnamefont {Strupinski}}, \bibinfo {author}
  {\bibfnamefont {Lyubov~G.}\ \bibnamefont {Bulusheva}}, \bibinfo {author}
  {\bibfnamefont {Alexander~V.}\ \bibnamefont {Okotrub}}, \bibinfo {author}
  {\bibfnamefont {Irina~V.}\ \bibnamefont {Grigorieva}}, \bibinfo {author}
  {\bibfnamefont {Alexander~N.}\ \bibnamefont {Grigorenko}}, \bibinfo {author}
  {\bibfnamefont {Kostya~S.}\ \bibnamefont {Novoselov}}, \ and\ \bibinfo
  {author} {\bibfnamefont {Andre~K.}\ \bibnamefont {Geim}},\ }\bibfield
  {title} {\enquote {\bibinfo {title} {Fluorographene: A two-dimensional
  counterpart of teflon},}\ }\href {\doibase 10.1002/smll.201001555} {\bibfield
   {journal} {\bibinfo  {journal} {Small}\ }\textbf {\bibinfo {volume} {6}},\
  \bibinfo {pages} {2877--2884} (\bibinfo {year} {2010})}\BibitemShut {NoStop}%
\bibitem [{\citenamefont {Zbořil}\ \emph {et~al.}(2010)\citenamefont
  {Zbořil}, \citenamefont {Karlický}, \citenamefont {Bourlinos},
  \citenamefont {Steriotis}, \citenamefont {Stubos}, \citenamefont
  {Georgakilas}, \citenamefont {Šafářová}, \citenamefont {Jančík},
  \citenamefont {Trapalis},\ and\ \citenamefont {Otyepka}}]{radek2010}%
  \BibitemOpen
  \bibfield  {author} {\bibinfo {author} {\bibfnamefont {Radek}\ \bibnamefont
  {Zbořil}}, \bibinfo {author} {\bibfnamefont {František}\ \bibnamefont
  {Karlický}}, \bibinfo {author} {\bibfnamefont {Athanasios~B.}\ \bibnamefont
  {Bourlinos}}, \bibinfo {author} {\bibfnamefont {Theodore~A.}\ \bibnamefont
  {Steriotis}}, \bibinfo {author} {\bibfnamefont {Athanasios~K.}\ \bibnamefont
  {Stubos}}, \bibinfo {author} {\bibfnamefont {Vasilios}\ \bibnamefont
  {Georgakilas}}, \bibinfo {author} {\bibfnamefont {Klára}\ \bibnamefont
  {Šafářová}}, \bibinfo {author} {\bibfnamefont {Dalibor}\ \bibnamefont
  {Jančík}}, \bibinfo {author} {\bibfnamefont {Christos}\ \bibnamefont
  {Trapalis}}, \ and\ \bibinfo {author} {\bibfnamefont {Michal}\ \bibnamefont
  {Otyepka}},\ }\bibfield  {title} {\enquote {\bibinfo {title} {Graphene
  fluoride: A stable stoichiometric graphene derivative and its chemical
  conversion to graphene},}\ }\href {\doibase 10.1002/smll.201001401}
  {\bibfield  {journal} {\bibinfo  {journal} {Small}\ }\textbf {\bibinfo
  {volume} {6}},\ \bibinfo {pages} {2885--2891} (\bibinfo {year}
  {2010})}\BibitemShut {NoStop}%
\bibitem [{\citenamefont {Withers}\ \emph {et~al.}(2010)\citenamefont
  {Withers}, \citenamefont {Dubois},\ and\ \citenamefont
  {Savchenko}}]{withers2010}%
  \BibitemOpen
  \bibfield  {author} {\bibinfo {author} {\bibfnamefont {F.}~\bibnamefont
  {Withers}}, \bibinfo {author} {\bibfnamefont {M.}~\bibnamefont {Dubois}}, \
  and\ \bibinfo {author} {\bibfnamefont {A.~K.}\ \bibnamefont {Savchenko}},\
  }\bibfield  {title} {\enquote {\bibinfo {title} {Electron properties of
  fluorinated single-layer graphene transistors},}\ }\href {\doibase
  10.1103/PhysRevB.82.073403} {\bibfield  {journal} {\bibinfo  {journal} {Phys.
  Rev. B}\ }\textbf {\bibinfo {volume} {82}},\ \bibinfo {pages} {073403}
  (\bibinfo {year} {2010})}\BibitemShut {NoStop}%
\bibitem [{\citenamefont {Li}\ \emph {et~al.}(2011)\citenamefont {Li},
  \citenamefont {Zhou}, \citenamefont {Wu}, \citenamefont {Peng}, \citenamefont
  {Yan}, \citenamefont {Zhou},\ and\ \citenamefont {Liu}}]{li2011}%
  \BibitemOpen
  \bibfield  {author} {\bibinfo {author} {\bibfnamefont {Bo}~\bibnamefont
  {Li}}, \bibinfo {author} {\bibfnamefont {Lin}\ \bibnamefont {Zhou}}, \bibinfo
  {author} {\bibfnamefont {Di}~\bibnamefont {Wu}}, \bibinfo {author}
  {\bibfnamefont {Hailin}\ \bibnamefont {Peng}}, \bibinfo {author}
  {\bibfnamefont {Kai}\ \bibnamefont {Yan}}, \bibinfo {author} {\bibfnamefont
  {Yu}~\bibnamefont {Zhou}}, \ and\ \bibinfo {author} {\bibfnamefont
  {Zhongfan}\ \bibnamefont {Liu}},\ }\bibfield  {title} {\enquote {\bibinfo
  {title} {Photochemical chlorination of graphene},}\ }\href {\doibase
  10.1021/nn201731t} {\bibfield  {journal} {\bibinfo  {journal} {ACS Nano}\
  }\textbf {\bibinfo {volume} {5}},\ \bibinfo {pages} {5957--5961} (\bibinfo
  {year} {2011})}\BibitemShut {NoStop}%
\bibitem [{\citenamefont {Garcia}\ \emph {et~al.}(2011)\citenamefont {Garcia},
  \citenamefont {de~Lima}, \citenamefont {Assali},\ and\ \citenamefont
  {Justo}}]{garcia2011}%
  \BibitemOpen
  \bibfield  {author} {\bibinfo {author} {\bibfnamefont {Joelson~C.}\
  \bibnamefont {Garcia}}, \bibinfo {author} {\bibfnamefont {Denille~B.}\
  \bibnamefont {de~Lima}}, \bibinfo {author} {\bibfnamefont {Lucy V.~C.}\
  \bibnamefont {Assali}}, \ and\ \bibinfo {author} {\bibfnamefont {João~F.}\
  \bibnamefont {Justo}},\ }\bibfield  {title} {\enquote {\bibinfo {title}
  {Group iv graphene- and graphane-like nanosheets},}\ }\href {\doibase
  10.1021/jp203657w} {\bibfield  {journal} {\bibinfo  {journal} {The Journal of
  Physical Chemistry C}\ }\textbf {\bibinfo {volume} {115}},\ \bibinfo {pages}
  {13242--13246} (\bibinfo {year} {2011})}\BibitemShut {NoStop}%
\bibitem [{\citenamefont {Sahin}\ and\ \citenamefont
  {Ciraci}(2012)}]{sahin2012}%
  \BibitemOpen
  \bibfield  {author} {\bibinfo {author} {\bibfnamefont {H.}~\bibnamefont
  {Sahin}}\ and\ \bibinfo {author} {\bibfnamefont {S.}~\bibnamefont {Ciraci}},\
  }\bibfield  {title} {\enquote {\bibinfo {title} {Chlorine adsorption on
  graphene: Chlorographene},}\ }\href {\doibase 10.1021/jp307006c} {\bibfield
  {journal} {\bibinfo  {journal} {The Journal of Physical Chemistry C}\
  }\textbf {\bibinfo {volume} {116}},\ \bibinfo {pages} {24075--24083}
  (\bibinfo {year} {2012})}\BibitemShut {NoStop}%
\bibitem [{\citenamefont {Conley}\ \emph {et~al.}(2013)\citenamefont {Conley},
  \citenamefont {Wang}, \citenamefont {Ziegler}, \citenamefont {Haglund},
  \citenamefont {Pantelides},\ and\ \citenamefont {Bolotin}}]{conley2013}%
  \BibitemOpen
  \bibfield  {author} {\bibinfo {author} {\bibfnamefont {Hiram~J.}\
  \bibnamefont {Conley}}, \bibinfo {author} {\bibfnamefont {Bin}\ \bibnamefont
  {Wang}}, \bibinfo {author} {\bibfnamefont {Jed~I.}\ \bibnamefont {Ziegler}},
  \bibinfo {author} {\bibfnamefont {Richard~F.}\ \bibnamefont {Haglund}},
  \bibinfo {author} {\bibfnamefont {Sokrates~T.}\ \bibnamefont {Pantelides}}, \
  and\ \bibinfo {author} {\bibfnamefont {Kirill~I.}\ \bibnamefont {Bolotin}},\
  }\bibfield  {title} {\enquote {\bibinfo {title} {Bandgap engineering of
  strained monolayer and bilayer mos2},}\ }\href {\doibase 10.1021/nl4014748}
  {\bibfield  {journal} {\bibinfo  {journal} {Nano Letters}\ }\textbf {\bibinfo
  {volume} {13}},\ \bibinfo {pages} {3626--3630} (\bibinfo {year}
  {2013})}\BibitemShut {NoStop}%
\bibitem [{\citenamefont {Lloyd}\ \emph {et~al.}(2016)\citenamefont {Lloyd},
  \citenamefont {Liu}, \citenamefont {Christopher}, \citenamefont {Cantley},
  \citenamefont {Wadehra}, \citenamefont {Kim}, \citenamefont {Goldberg},
  \citenamefont {Swan},\ and\ \citenamefont {Bunch}}]{lloyd2016}%
  \BibitemOpen
  \bibfield  {author} {\bibinfo {author} {\bibfnamefont {David}\ \bibnamefont
  {Lloyd}}, \bibinfo {author} {\bibfnamefont {Xinghui}\ \bibnamefont {Liu}},
  \bibinfo {author} {\bibfnamefont {Jason~W.}\ \bibnamefont {Christopher}},
  \bibinfo {author} {\bibfnamefont {Lauren}\ \bibnamefont {Cantley}}, \bibinfo
  {author} {\bibfnamefont {Anubhav}\ \bibnamefont {Wadehra}}, \bibinfo {author}
  {\bibfnamefont {Brian~L.}\ \bibnamefont {Kim}}, \bibinfo {author}
  {\bibfnamefont {Bennett~B.}\ \bibnamefont {Goldberg}}, \bibinfo {author}
  {\bibfnamefont {Anna~K.}\ \bibnamefont {Swan}}, \ and\ \bibinfo {author}
  {\bibfnamefont {J.~Scott}\ \bibnamefont {Bunch}},\ }\bibfield  {title}
  {\enquote {\bibinfo {title} {Band gap engineering with ultralarge biaxial
  strains in suspended monolayer mos2},}\ }\href {\doibase
  10.1021/acs.nanolett.6b02615} {\bibfield  {journal} {\bibinfo  {journal}
  {Nano Letters}\ }\textbf {\bibinfo {volume} {16}},\ \bibinfo {pages}
  {5836--5841} (\bibinfo {year} {2016})}\BibitemShut {NoStop}%
\bibitem [{\citenamefont {Zhu}\ and\ \citenamefont
  {Tom\'anek}(2014)}]{zhu2014}%
  \BibitemOpen
  \bibfield  {author} {\bibinfo {author} {\bibfnamefont {Zhen}\ \bibnamefont
  {Zhu}}\ and\ \bibinfo {author} {\bibfnamefont {David}\ \bibnamefont
  {Tom\'anek}},\ }\bibfield  {title} {\enquote {\bibinfo {title}
  {Semiconducting layered blue phosphorus: A computational study},}\ }\href
  {\doibase 10.1103/PhysRevLett.112.176802} {\bibfield  {journal} {\bibinfo
  {journal} {Phys. Rev. Lett.}\ }\textbf {\bibinfo {volume} {112}},\ \bibinfo
  {pages} {176802} (\bibinfo {year} {2014})}\BibitemShut {NoStop}%
\bibitem [{\citenamefont {Guan}\ \emph {et~al.}(2014)\citenamefont {Guan},
  \citenamefont {Zhu},\ and\ \citenamefont {Tom\'anek}}]{guan2014}%
  \BibitemOpen
  \bibfield  {author} {\bibinfo {author} {\bibfnamefont {Jie}\ \bibnamefont
  {Guan}}, \bibinfo {author} {\bibfnamefont {Zhen}\ \bibnamefont {Zhu}}, \ and\
  \bibinfo {author} {\bibfnamefont {David}\ \bibnamefont {Tom\'anek}},\
  }\bibfield  {title} {\enquote {\bibinfo {title} {Phase coexistence and
  metal-insulator transition in few-layer phosphorene: A computational
  study},}\ }\href {\doibase 10.1103/PhysRevLett.113.046804} {\bibfield
  {journal} {\bibinfo  {journal} {Phys. Rev. Lett.}\ }\textbf {\bibinfo
  {volume} {113}},\ \bibinfo {pages} {046804} (\bibinfo {year}
  {2014})}\BibitemShut {NoStop}%
\bibitem [{\citenamefont {Fei}\ and\ \citenamefont {Yang}(2014)}]{fei2014}%
  \BibitemOpen
  \bibfield  {author} {\bibinfo {author} {\bibfnamefont {Ruixiang}\
  \bibnamefont {Fei}}\ and\ \bibinfo {author} {\bibfnamefont {Li}~\bibnamefont
  {Yang}},\ }\bibfield  {title} {\enquote {\bibinfo {title} {Strain-engineering
  the anisotropic electrical conductance of few-layer black phosphorus},}\
  }\href {\doibase 10.1021/nl500935z} {\bibfield  {journal} {\bibinfo
  {journal} {Nano Letters}\ }\textbf {\bibinfo {volume} {14}},\ \bibinfo
  {pages} {2884--2889} (\bibinfo {year} {2014})}\BibitemShut {NoStop}%
\bibitem [{\citenamefont {Lu}\ \emph {et~al.}(2014)\citenamefont {Lu},
  \citenamefont {Guo}, \citenamefont {Li}, \citenamefont {Dai}, \citenamefont
  {Wang}, \citenamefont {Mei}, \citenamefont {Wu},\ and\ \citenamefont
  {Zeng}}]{lupccp2014}%
  \BibitemOpen
  \bibfield  {author} {\bibinfo {author} {\bibfnamefont {Ning}\ \bibnamefont
  {Lu}}, \bibinfo {author} {\bibfnamefont {Hongyan}\ \bibnamefont {Guo}},
  \bibinfo {author} {\bibfnamefont {Lei}\ \bibnamefont {Li}}, \bibinfo {author}
  {\bibfnamefont {Jun}\ \bibnamefont {Dai}}, \bibinfo {author} {\bibfnamefont
  {Lu}~\bibnamefont {Wang}}, \bibinfo {author} {\bibfnamefont {Wai-Ning}\
  \bibnamefont {Mei}}, \bibinfo {author} {\bibfnamefont {Xiaojun}\ \bibnamefont
  {Wu}}, \ and\ \bibinfo {author} {\bibfnamefont {Xiao~Cheng}\ \bibnamefont
  {Zeng}},\ }\bibfield  {title} {\enquote {\bibinfo {title} {Mos2/mx2
  heterobilayers: bandgap engineering via tensile strain or external electrical
  field},}\ }\href {\doibase 10.1039/C3NR06072A} {\bibfield  {journal}
  {\bibinfo  {journal} {Nanoscale}\ }\textbf {\bibinfo {volume} {6}},\ \bibinfo
  {pages} {2879--2886} (\bibinfo {year} {2014})}\BibitemShut {NoStop}%
\bibitem [{\citenamefont {Qi}\ \emph {et~al.}(2013)\citenamefont {Qi},
  \citenamefont {Li}, \citenamefont {Qian},\ and\ \citenamefont
  {Feng}}]{qiapl2013}%
  \BibitemOpen
  \bibfield  {author} {\bibinfo {author} {\bibfnamefont {Jingshan}\
  \bibnamefont {Qi}}, \bibinfo {author} {\bibfnamefont {Xiao}\ \bibnamefont
  {Li}}, \bibinfo {author} {\bibfnamefont {Xiaofeng}\ \bibnamefont {Qian}}, \
  and\ \bibinfo {author} {\bibfnamefont {Ji}~\bibnamefont {Feng}},\ }\bibfield
  {title} {\enquote {\bibinfo {title} {Bandgap engineering of rippled mos2
  monolayer under external electric field},}\ }\href {\doibase
  10.1063/1.4803803} {\bibfield  {journal} {\bibinfo  {journal} {Applied
  Physics Letters}\ }\textbf {\bibinfo {volume} {102}},\ \bibinfo {pages}
  {173112} (\bibinfo {year} {2013})}\BibitemShut {NoStop}%
\bibitem [{\citenamefont {Liu}\ \emph {et~al.}(2012)\citenamefont {Liu},
  \citenamefont {Li}, \citenamefont {Li}, \citenamefont {Gao}, \citenamefont
  {Chen},\ and\ \citenamefont {Lu}}]{liujpcc2012}%
  \BibitemOpen
  \bibfield  {author} {\bibinfo {author} {\bibfnamefont {Qihang}\ \bibnamefont
  {Liu}}, \bibinfo {author} {\bibfnamefont {Linze}\ \bibnamefont {Li}},
  \bibinfo {author} {\bibfnamefont {Yafei}\ \bibnamefont {Li}}, \bibinfo
  {author} {\bibfnamefont {Zhengxiang}\ \bibnamefont {Gao}}, \bibinfo {author}
  {\bibfnamefont {Zhongfang}\ \bibnamefont {Chen}}, \ and\ \bibinfo {author}
  {\bibfnamefont {Jing}\ \bibnamefont {Lu}},\ }\bibfield  {title} {\enquote
  {\bibinfo {title} {Tuning electronic structure of bilayer mos2 by vertical
  electric field: A first-principles investigation},}\ }\href {\doibase
  10.1021/jp307124d} {\bibfield  {journal} {\bibinfo  {journal} {The Journal of
  Physical Chemistry C}\ }\textbf {\bibinfo {volume} {116}},\ \bibinfo {pages}
  {21556--21562} (\bibinfo {year} {2012})}\BibitemShut {NoStop}%
\bibitem [{\citenamefont {Liu}\ \emph {et~al.}(2015)\citenamefont {Liu},
  \citenamefont {Zhang}, \citenamefont {Abdalla}, \citenamefont {Fazzio},\ and\
  \citenamefont {Zunger}}]{liunl2015}%
  \BibitemOpen
  \bibfield  {author} {\bibinfo {author} {\bibfnamefont {Qihang}\ \bibnamefont
  {Liu}}, \bibinfo {author} {\bibfnamefont {Xiuwen}\ \bibnamefont {Zhang}},
  \bibinfo {author} {\bibfnamefont {L.~B.}\ \bibnamefont {Abdalla}}, \bibinfo
  {author} {\bibfnamefont {Adalberto}\ \bibnamefont {Fazzio}}, \ and\ \bibinfo
  {author} {\bibfnamefont {Alex}\ \bibnamefont {Zunger}},\ }\bibfield  {title}
  {\enquote {\bibinfo {title} {Switching a normal insulator into a topological
  insulator via electric field with application to phosphorene},}\ }\href
  {\doibase 10.1021/nl5043769} {\bibfield  {journal} {\bibinfo  {journal} {Nano
  Letters}\ }\textbf {\bibinfo {volume} {15}},\ \bibinfo {pages} {1222--1228}
  (\bibinfo {year} {2015})}\BibitemShut {NoStop}%
\bibitem [{\citenamefont {Dai}\ and\ \citenamefont {Zeng}(2014)}]{dai2014}%
  \BibitemOpen
  \bibfield  {author} {\bibinfo {author} {\bibfnamefont {Jun}\ \bibnamefont
  {Dai}}\ and\ \bibinfo {author} {\bibfnamefont {Xiao~Cheng}\ \bibnamefont
  {Zeng}},\ }\bibfield  {title} {\enquote {\bibinfo {title} {Bilayer
  phosphorene: Effect of stacking order on bandgap and its potential
  applications in thin-film solar cells},}\ }\href {\doibase 10.1021/jz500409m}
  {\bibfield  {journal} {\bibinfo  {journal} {The Journal of Physical Chemistry
  Letters}\ }\textbf {\bibinfo {volume} {5}},\ \bibinfo {pages} {1289--1293}
  (\bibinfo {year} {2014})}\BibitemShut {NoStop}%
\bibitem [{\citenamefont {Li}\ \emph {et~al.}(2017)\citenamefont {Li},
  \citenamefont {Xu}, \citenamefont {Ba}, \citenamefont {Xuan}, \citenamefont
  {Chen}, \citenamefont {Sun}, \citenamefont {Zhang},\ and\ \citenamefont
  {Zhang}}]{dong2017}%
  \BibitemOpen
  \bibfield  {author} {\bibinfo {author} {\bibfnamefont {Dong}\ \bibnamefont
  {Li}}, \bibinfo {author} {\bibfnamefont {Jin-Rong}\ \bibnamefont {Xu}},
  \bibinfo {author} {\bibfnamefont {Kun}\ \bibnamefont {Ba}}, \bibinfo {author}
  {\bibfnamefont {Ningning}\ \bibnamefont {Xuan}}, \bibinfo {author}
  {\bibfnamefont {Mingyuan}\ \bibnamefont {Chen}}, \bibinfo {author}
  {\bibfnamefont {Zhengzong}\ \bibnamefont {Sun}}, \bibinfo {author}
  {\bibfnamefont {Yu-Zhong}\ \bibnamefont {Zhang}}, \ and\ \bibinfo {author}
  {\bibfnamefont {Zengxing}\ \bibnamefont {Zhang}},\ }\bibfield  {title}
  {\enquote {\bibinfo {title} {Tunable bandgap in few-layer black phosphorus by
  electrical field},}\ }\href {http://stacks.iop.org/2053-1583/4/i=3/a=031009}
  {\bibfield  {journal} {\bibinfo  {journal} {2D Materials}\ }\textbf {\bibinfo
  {volume} {4}},\ \bibinfo {pages} {031009} (\bibinfo {year}
  {2017})}\BibitemShut {NoStop}%
\bibitem [{\citenamefont {Ghosh}\ \emph {et~al.}(2015)\citenamefont {Ghosh},
  \citenamefont {Nahas}, \citenamefont {Bhowmick},\ and\ \citenamefont
  {Agarwal}}]{ghosh2015}%
  \BibitemOpen
  \bibfield  {author} {\bibinfo {author} {\bibfnamefont {Barun}\ \bibnamefont
  {Ghosh}}, \bibinfo {author} {\bibfnamefont {Suhas}\ \bibnamefont {Nahas}},
  \bibinfo {author} {\bibfnamefont {Somnath}\ \bibnamefont {Bhowmick}}, \ and\
  \bibinfo {author} {\bibfnamefont {Amit}\ \bibnamefont {Agarwal}},\ }\bibfield
   {title} {\enquote {\bibinfo {title} {Electric field induced gap modification
  in ultrathin blue phosphorus},}\ }\href {\doibase 10.1103/PhysRevB.91.115433}
  {\bibfield  {journal} {\bibinfo  {journal} {Phys. Rev. B}\ }\textbf {\bibinfo
  {volume} {91}},\ \bibinfo {pages} {115433} (\bibinfo {year}
  {2015})}\BibitemShut {NoStop}%
\bibitem [{\citenamefont {Tang}\ \emph {et~al.}(2011)\citenamefont {Tang},
  \citenamefont {Qin}, \citenamefont {Zhou}, \citenamefont {Qu}, \citenamefont
  {Zheng}, \citenamefont {Fei}, \citenamefont {Li}, \citenamefont {Zheng},
  \citenamefont {Gao},\ and\ \citenamefont {Lu}}]{tang2011}%
  \BibitemOpen
  \bibfield  {author} {\bibinfo {author} {\bibfnamefont {Kechao}\ \bibnamefont
  {Tang}}, \bibinfo {author} {\bibfnamefont {Rui}\ \bibnamefont {Qin}},
  \bibinfo {author} {\bibfnamefont {Jing}\ \bibnamefont {Zhou}}, \bibinfo
  {author} {\bibfnamefont {Heruge}\ \bibnamefont {Qu}}, \bibinfo {author}
  {\bibfnamefont {Jiaxin}\ \bibnamefont {Zheng}}, \bibinfo {author}
  {\bibfnamefont {Ruixiang}\ \bibnamefont {Fei}}, \bibinfo {author}
  {\bibfnamefont {Hong}\ \bibnamefont {Li}}, \bibinfo {author} {\bibfnamefont
  {Qiye}\ \bibnamefont {Zheng}}, \bibinfo {author} {\bibfnamefont {Zhengxiang}\
  \bibnamefont {Gao}}, \ and\ \bibinfo {author} {\bibfnamefont {Jing}\
  \bibnamefont {Lu}},\ }\bibfield  {title} {\enquote {\bibinfo {title}
  {Electric-field-induced energy gap in few-layer graphene},}\ }\href {\doibase
  10.1021/jp201761p} {\bibfield  {journal} {\bibinfo  {journal} {The Journal of
  Physical Chemistry C}\ }\textbf {\bibinfo {volume} {115}},\ \bibinfo {pages}
  {9458--9464} (\bibinfo {year} {2011})}\BibitemShut {NoStop}%
\bibitem [{\citenamefont {Kokalj}(2003)}]{Kokalj03}%
  \BibitemOpen
  \bibfield  {author} {\bibinfo {author} {\bibfnamefont {Anton}\ \bibnamefont
  {Kokalj}},\ }\bibfield  {title} {\enquote {\bibinfo {title} {Computer
  graphics and graphical user interfaces as tools in simulations of matter at
  the atomic scale},}\ }\href {\doibase
  http://dx.doi.org/10.1016/S0927-0256(03)00104-6} {\bibfield  {journal}
  {\bibinfo  {journal} {Computational Materials Science}\ }\textbf {\bibinfo
  {volume} {28}},\ \bibinfo {pages} {155 -- 168} (\bibinfo {year} {2003})},\
  \bibinfo {note} {proceedings of the Symposium on Software Development for
  Process and Materials Design}\BibitemShut {NoStop}%
\bibitem [{\citenamefont {Kokalj}(1999)}]{Kokalj99}%
  \BibitemOpen
  \bibfield  {author} {\bibinfo {author} {\bibfnamefont {Anton}\ \bibnamefont
  {Kokalj}},\ }\bibfield  {title} {\enquote {\bibinfo {title} {Xcrysden—a new
  program for displaying crystalline structures and electron densities},}\
  }\href {\doibase http://dx.doi.org/10.1016/S1093-3263(99)00028-5} {\bibfield
  {journal} {\bibinfo  {journal} {Journal of Molecular Graphics and Modelling}\
  }\textbf {\bibinfo {volume} {17}},\ \bibinfo {pages} {176 -- 179} (\bibinfo
  {year} {1999})}\BibitemShut {NoStop}%
\bibitem [{\citenamefont {Giannozzi}\ \emph {et~al.}(2009)\citenamefont
  {Giannozzi}, \citenamefont {Baroni}, \citenamefont {Bonini}, \citenamefont
  {Calandra}, \citenamefont {Car}, \citenamefont {Cavazzoni}, \citenamefont
  {Ceresoli}, \citenamefont {Chiarotti}, \citenamefont {Cococcioni},
  \citenamefont {Dabo}, \citenamefont {{Dal Corso}}, \citenamefont
  {de~Gironcoli}, \citenamefont {Fabris}, \citenamefont {Fratesi},
  \citenamefont {Gebauer}, \citenamefont {Gerstmann}, \citenamefont
  {Gougoussis}, \citenamefont {Kokalj}, \citenamefont {Lazzeri}, \citenamefont
  {Martin-Samos}, \citenamefont {Marzari}, \citenamefont {Mauri}, \citenamefont
  {Mazzarello}, \citenamefont {Paolini}, \citenamefont {Pasquarello},
  \citenamefont {Paulatto}, \citenamefont {Sbraccia}, \citenamefont {Scandolo},
  \citenamefont {Sclauzero}, \citenamefont {Seitsonen}, \citenamefont
  {Smogunov}, \citenamefont {Umari},\ and\ \citenamefont
  {Wentzcovitch}}]{Gianozzi09}%
  \BibitemOpen
  \bibfield  {author} {\bibinfo {author} {\bibfnamefont {Paolo}\ \bibnamefont
  {Giannozzi}}, \bibinfo {author} {\bibfnamefont {Stefano}\ \bibnamefont
  {Baroni}}, \bibinfo {author} {\bibfnamefont {Nicola}\ \bibnamefont {Bonini}},
  \bibinfo {author} {\bibfnamefont {Matteo}\ \bibnamefont {Calandra}}, \bibinfo
  {author} {\bibfnamefont {Roberto}\ \bibnamefont {Car}}, \bibinfo {author}
  {\bibfnamefont {Carlo}\ \bibnamefont {Cavazzoni}}, \bibinfo {author}
  {\bibfnamefont {Davide}\ \bibnamefont {Ceresoli}}, \bibinfo {author}
  {\bibfnamefont {Guido~L}\ \bibnamefont {Chiarotti}}, \bibinfo {author}
  {\bibfnamefont {Matteo}\ \bibnamefont {Cococcioni}}, \bibinfo {author}
  {\bibfnamefont {Ismaila}\ \bibnamefont {Dabo}}, \bibinfo {author}
  {\bibfnamefont {Andrea}\ \bibnamefont {{Dal Corso}}}, \bibinfo {author}
  {\bibfnamefont {Stefano}\ \bibnamefont {de~Gironcoli}}, \bibinfo {author}
  {\bibfnamefont {Stefano}\ \bibnamefont {Fabris}}, \bibinfo {author}
  {\bibfnamefont {Guido}\ \bibnamefont {Fratesi}}, \bibinfo {author}
  {\bibfnamefont {Ralph}\ \bibnamefont {Gebauer}}, \bibinfo {author}
  {\bibfnamefont {Uwe}\ \bibnamefont {Gerstmann}}, \bibinfo {author}
  {\bibfnamefont {Christos}\ \bibnamefont {Gougoussis}}, \bibinfo {author}
  {\bibfnamefont {Anton}\ \bibnamefont {Kokalj}}, \bibinfo {author}
  {\bibfnamefont {Michele}\ \bibnamefont {Lazzeri}}, \bibinfo {author}
  {\bibfnamefont {Layla}\ \bibnamefont {Martin-Samos}}, \bibinfo {author}
  {\bibfnamefont {Nicola}\ \bibnamefont {Marzari}}, \bibinfo {author}
  {\bibfnamefont {Francesco}\ \bibnamefont {Mauri}}, \bibinfo {author}
  {\bibfnamefont {Riccardo}\ \bibnamefont {Mazzarello}}, \bibinfo {author}
  {\bibfnamefont {Stefano}\ \bibnamefont {Paolini}}, \bibinfo {author}
  {\bibfnamefont {Alfredo}\ \bibnamefont {Pasquarello}}, \bibinfo {author}
  {\bibfnamefont {Lorenzo}\ \bibnamefont {Paulatto}}, \bibinfo {author}
  {\bibfnamefont {Carlo}\ \bibnamefont {Sbraccia}}, \bibinfo {author}
  {\bibfnamefont {Sandro}\ \bibnamefont {Scandolo}}, \bibinfo {author}
  {\bibfnamefont {Gabriele}\ \bibnamefont {Sclauzero}}, \bibinfo {author}
  {\bibfnamefont {Ari~P}\ \bibnamefont {Seitsonen}}, \bibinfo {author}
  {\bibfnamefont {Alexander}\ \bibnamefont {Smogunov}}, \bibinfo {author}
  {\bibfnamefont {Paolo}\ \bibnamefont {Umari}}, \ and\ \bibinfo {author}
  {\bibfnamefont {Renata~M}\ \bibnamefont {Wentzcovitch}},\ }\bibfield  {title}
  {\enquote {\bibinfo {title} {Quantum espresso: a modular and open-source
  software project for quantum simulations of materials},}\ }\href
  {http://www.quantum-espresso.org} {\bibfield  {journal} {\bibinfo  {journal}
  {Journal of Physics: Condensed Matter}\ }\textbf {\bibinfo {volume} {21}},\
  \bibinfo {pages} {395502 (19pp)} (\bibinfo {year} {2009})}\BibitemShut
  {NoStop}%
\bibitem [{Note1()}]{Note1}%
  \BibitemOpen
  \bibinfo {note} {Point group symmetry of the wave vector $\protect \bm {k}$
  at the $\Gamma $, K and M point is D$_{3d}$, D$_3$ and C$_{2h}$, respectively
  and along the $\Gamma $-K, K-M and M-$\Gamma $ line is C$_2$, C$_2$ and
  C$_s$, respectively.}\BibitemShut {Stop}%
\bibitem [{\citenamefont {Mardanya}\ \emph {et~al.}(2016)\citenamefont
  {Mardanya}, \citenamefont {Thakur}, \citenamefont {Bhowmick},\ and\
  \citenamefont {Agarwal}}]{Mardanya:2016}%
  \BibitemOpen
  \bibfield  {author} {\bibinfo {author} {\bibfnamefont {Sougata}\ \bibnamefont
  {Mardanya}}, \bibinfo {author} {\bibfnamefont {Vinay~Kumar}\ \bibnamefont
  {Thakur}}, \bibinfo {author} {\bibfnamefont {Somnath}\ \bibnamefont
  {Bhowmick}}, \ and\ \bibinfo {author} {\bibfnamefont {Amit}\ \bibnamefont
  {Agarwal}},\ }\bibfield  {title} {\enquote {\bibinfo {title} {Four allotropes
  of semiconducting layered arsenic that switch into a topological insulator
  via an ele ctric field: Computational study},}\ }\href {\doibase
  10.1103/PhysRevB.94.035423} {\bibfield  {journal} {\bibinfo  {journal} {Phys.
  Rev. B}\ }\textbf {\bibinfo {volume} {94}},\ \bibinfo {pages} {035423}
  (\bibinfo {year} {2016})}\BibitemShut {NoStop}%
\bibitem [{\citenamefont {Zhu}\ \emph {et~al.}(2016)\citenamefont {Zhu},
  \citenamefont {Wang}, \citenamefont {Guan}, \citenamefont {Liu},
  \citenamefont {Zhang}, \citenamefont {Chen},\ and\ \citenamefont
  {Yang}}]{zhu2016}%
  \BibitemOpen
  \bibfield  {author} {\bibinfo {author} {\bibfnamefont {Liyan}\ \bibnamefont
  {Zhu}}, \bibinfo {author} {\bibfnamefont {Shan-Shan}\ \bibnamefont {Wang}},
  \bibinfo {author} {\bibfnamefont {Shan}\ \bibnamefont {Guan}}, \bibinfo
  {author} {\bibfnamefont {Ying}\ \bibnamefont {Liu}}, \bibinfo {author}
  {\bibfnamefont {Tingting}\ \bibnamefont {Zhang}}, \bibinfo {author}
  {\bibfnamefont {Guibin}\ \bibnamefont {Chen}}, \ and\ \bibinfo {author}
  {\bibfnamefont {Shengyuan~A.}\ \bibnamefont {Yang}},\ }\bibfield  {title}
  {\enquote {\bibinfo {title} {Blue phosphorene oxide: Strain-tunable quantum
  phase transitions and novel 2d emergent fermions},}\ }\href {\doibase
  10.1021/acs.nanolett.6b03208} {\bibfield  {journal} {\bibinfo  {journal}
  {Nano Letters}\ }\textbf {\bibinfo {volume} {16}},\ \bibinfo {pages}
  {6548--6554} (\bibinfo {year} {2016})}\BibitemShut {NoStop}%
\bibitem [{\citenamefont {Konschuh}\ \emph {et~al.}(2010)\citenamefont
  {Konschuh}, \citenamefont {Gmitra},\ and\ \citenamefont
  {Fabian}}]{sergej2010}%
  \BibitemOpen
  \bibfield  {author} {\bibinfo {author} {\bibfnamefont {Sergej}\ \bibnamefont
  {Konschuh}}, \bibinfo {author} {\bibfnamefont {Martin}\ \bibnamefont
  {Gmitra}}, \ and\ \bibinfo {author} {\bibfnamefont {Jaroslav}\ \bibnamefont
  {Fabian}},\ }\bibfield  {title} {\enquote {\bibinfo {title} {Tight-binding
  theory of the spin-orbit coupling in graphene},}\ }\href {\doibase
  10.1103/PhysRevB.82.245412} {\bibfield  {journal} {\bibinfo  {journal} {Phys.
  Rev. B}\ }\textbf {\bibinfo {volume} {82}},\ \bibinfo {pages} {245412}
  (\bibinfo {year} {2010})}\BibitemShut {NoStop}%
\bibitem [{Note2()}]{Note2}%
  \BibitemOpen
  \bibinfo {note} {Fully relativistic pseudopotentials are used for spin-orbit
  coupling calculations.}\BibitemShut {Stop}%
\bibitem [{\citenamefont {Fu}\ and\ \citenamefont {Kane}(2007)}]{Fu2007}%
  \BibitemOpen
  \bibfield  {author} {\bibinfo {author} {\bibfnamefont {Liang}\ \bibnamefont
  {Fu}}\ and\ \bibinfo {author} {\bibfnamefont {C.~L.}\ \bibnamefont {Kane}},\
  }\bibfield  {title} {\enquote {\bibinfo {title} {Topological insulators with
  inversion symmetry},}\ }\href {\doibase 10.1103/PhysRevB.76.045302}
  {\bibfield  {journal} {\bibinfo  {journal} {Phys. Rev. B}\ }\textbf {\bibinfo
  {volume} {76}},\ \bibinfo {pages} {045302} (\bibinfo {year}
  {2007})}\BibitemShut {NoStop}%
\bibitem [{\citenamefont {Karlický}\ and\ \citenamefont
  {Otyepka}(2013)}]{jctc2013}%
  \BibitemOpen
  \bibfield  {author} {\bibinfo {author} {\bibfnamefont {František}\
  \bibnamefont {Karlický}}\ and\ \bibinfo {author} {\bibfnamefont {Michal}\
  \bibnamefont {Otyepka}},\ }\bibfield  {title} {\enquote {\bibinfo {title}
  {Band gaps and optical spectra of chlorographene, fluorographene and graphane
  from g0w0, gw0 and gw calculations on top of pbe and hse06 orbitals},}\
  }\href {\doibase 10.1021/ct400476r} {\bibfield  {journal} {\bibinfo
  {journal} {Journal of Chemical Theory and Computation}\ }\textbf {\bibinfo
  {volume} {9}},\ \bibinfo {pages} {4155--4164} (\bibinfo {year}
  {2013})}\BibitemShut {NoStop}%
\bibitem [{\citenamefont {Crowley}\ \emph {et~al.}(2016)\citenamefont
  {Crowley}, \citenamefont {Tahir-Kheli},\ and\ \citenamefont
  {Goddard}}]{crowley2016}%
  \BibitemOpen
  \bibfield  {author} {\bibinfo {author} {\bibfnamefont {Jason~M.}\
  \bibnamefont {Crowley}}, \bibinfo {author} {\bibfnamefont {Jamil}\
  \bibnamefont {Tahir-Kheli}}, \ and\ \bibinfo {author} {\bibfnamefont
  {William~A.}\ \bibnamefont {Goddard}},\ }\bibfield  {title} {\enquote
  {\bibinfo {title} {Resolution of the band gap prediction problem for
  materials design},}\ }\href {\doibase 10.1021/acs.jpclett.5b02870} {\bibfield
   {journal} {\bibinfo  {journal} {The Journal of Physical Chemistry Letters}\
  }\textbf {\bibinfo {volume} {7}},\ \bibinfo {pages} {1198--1203} (\bibinfo
  {year} {2016})}\BibitemShut {NoStop}%
\end{thebibliography}%
\end{document}